\documentclass[aps,prl,floats,twocolumn,floatfix,showpacs]{revtex4}
\pdfoutput=1
\usepackage{epsf,graphicx}
\usepackage{amssymb}
\usepackage{amsmath}

\begin{document}

\title{Correlated electron tunneling through two separate quantum dot systems with strong capacitive interdot coupling}

\author{A.\ H\"ubel$^1$, K.\ Held$^{1,2}$, J.\ Weis$^1$, and K.\ v.\ Klitzing$^1$}
\affiliation{$^1$ Max-Planck-Institut f\"ur Festk\"orperforschung,
Heisenbergstr.~1, D-70569 Stuttgart, Germany}
\affiliation{$^2$ Institute for Solid State Physics, Vienna University of Technology,
1040 Vienna, Austria}

\date{\today}

\begin{abstract}
A system consisting of two independently contacted quantum dots with
strong electrostatic interaction shows interdot Coulomb blockade
when the dots are weakly tunnel coupled to their leads. It is
studied experimentally how the blockade can be overcome by
correlated tunneling when tunnel coupling to the leads increases.
The experimental results are compared with numerical renormalization
group  calculations using predefined (measured) parameters. Our results
indicate Kondo correlations due to the electrostatic interaction in
this double quantum dot system.
\end{abstract}

\maketitle

Electrical transport through a quantum dot at low temperature is
dominated by the electron-electron interaction, leading to Coulomb
blockade and single-electron charging effects \cite{Beenakker1}. The
spin as an internal degree of freedom causes under certain
circumstances a Kondo correlated state to form between the quantum dot
and its source and drain leads, overcoming the Coulomb blockade
with decreasing temperature \cite{Goldhaber-Gordon1}. The Anderson
impurity model not only provides a simplified, yet appropriate
description for this particular effect but, at the same time, it is the basic
model for a quantum dot system, i.e., a single localized orbital
tunnel coupled to leads \cite{Glazman1}. Two independently contacted
quantum dots with purely capacitive interaction can be labeled by a
pseudo-spin index and can therefore be described as another
realization of the Anderson impurity model \cite{Wilhelm1}. Theory
predicts that correlations should lift the Coulomb blockade where an
electrostatic degeneracy exists between states with  $(N_1, N_2)$
and $(N_1\pm 1,N_2\mp 1)$ electrons on the two dots. Experimentally
one would observe the Kondo correlations under such degeneracy
conditions upon enhancing the tunnel couplings of the dots to the
leads or lowering the temperature. When the spin is included, Kondo physics
with SU(4) symmetry can be present \cite{Borda1}. Experimental
results on cylindrical quantum dots \cite{Sasaki1} and carbon
nanotubes \cite{CNT} have been interpreted in terms of an SU(4)
spin-orbital Kondo effect. However the tunneling paths via the two
orbitals were not separately accessible to experiment, and therefore
assumptions about them had to be made. In contrast, the setup of
separate quantum dot systems with interdot capacitive coupling
allows one to study the (pseudo)spin-polarized currents and
therefore the Kondo correlations in a controlled way, provided the
conductances through the two quantum dots can be monitored
independently and for different parameter combinations. Several
experiments have examined the behavior of such samples at weak
tunnel couplings, where single-electron tunneling is an appropriate
description \cite{Chan1}, showing the expected honeycomb-like charge
stability diagram with less pronounced capacitive interdot coupling.
Two \emph{vertically} stacked quantum dot
systems show strong interdot capacitive coupling, and indications of
Kondo correlations have been observed, however the structure lacks
full control over the tunnel couplings \cite{Wilhelm2}.

\par
In this letter we use a double quantum dot system in lateral
arrangement with strong capacitive interdot interaction and fully
tunable tunnel couplings \cite{Huebel1}. The regions where transport
is dominated by interdot correlations are well-resolved, and we
study experimentally the transition from weak to strong tunnel
coupling. The conductances and the parameters of both quantum dots
are measured independently, so we can directly compare with
numerical renormalization group (NRG) calculations identifying
Kondoesque correlations.

\par
\begin{figure}[h]
    \includegraphics[angle=-90,width=8cm]
      {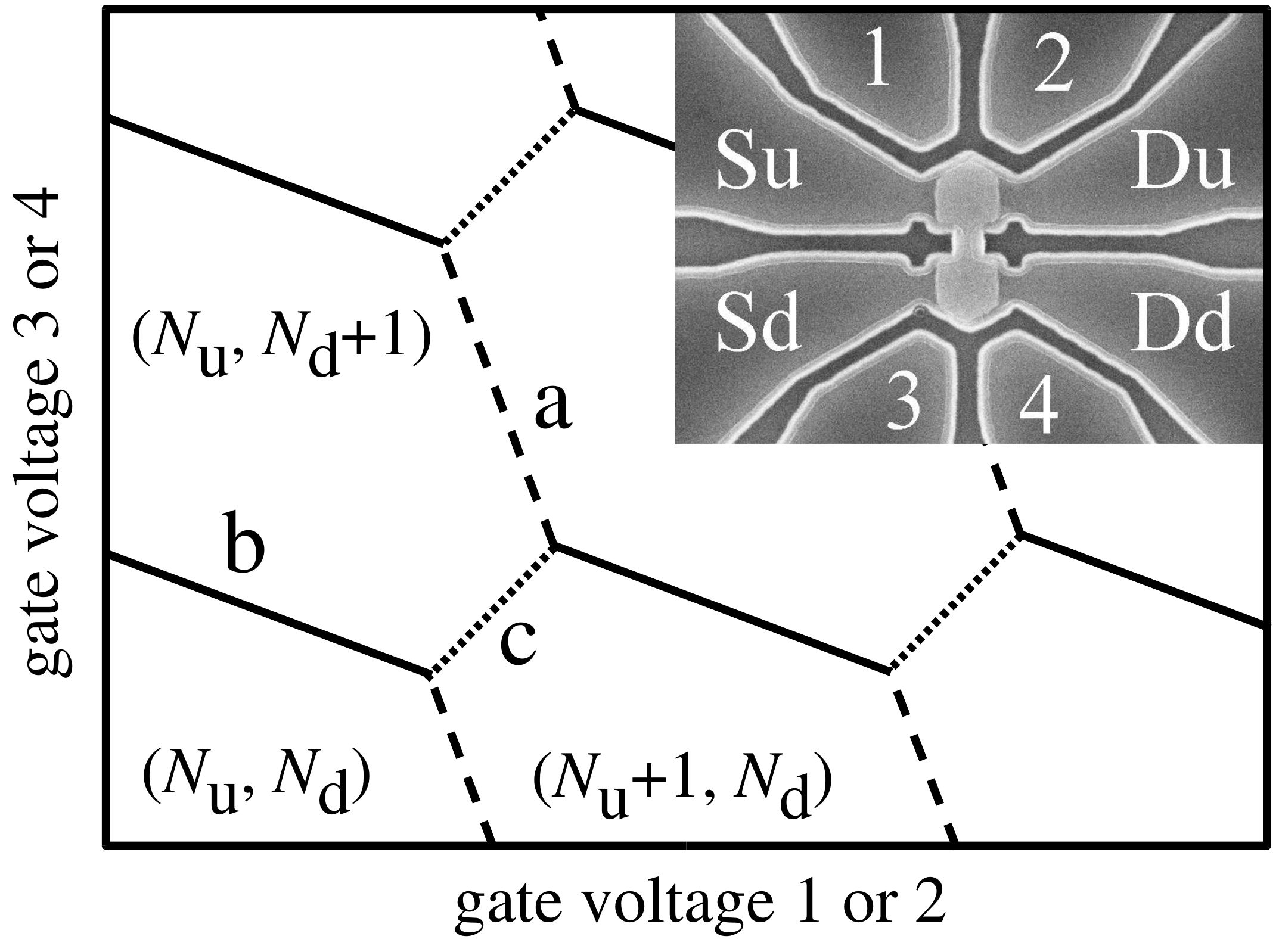}
    \caption{\label{figure1}
    Scheme of charge stability in a double quantum dot system as a
    function of two gate voltages, reflecting the characteristic honeycomb-like structure.
    At weak tunnel couplings, single-electron transport is possible through one of the dots on
    hexagon edges of type $a$ and $b$, respectively. Correlations between the
    quantum dots can lead to transport on type $c$ edges for enhanced tunnel couplings.
    Inset: SEM image with labeled electrodes and floating top gate in the center.}
\end{figure}

A scanning electron microscope (SEM) image of our sample is shown in
the inset of Fig.~\ref{figure1}. Its design concept and its
fabrication have been described elsewhere in more detail
\cite{Huebel1}. We use a GaAs/Al$_{0.33}$Ga$_{0.67}$As
heterostructure with a two-dimensional electron system, located at
the heterojunction 50 nm below the surface (electron density:
$3.2\times 10^{11}$ cm$^{-2}$, mobility: $3.0\times 10^5$
cm$^2$V$^{-1}$s$^{-1}$ at 4.2 K). First, we define a floating
metallic top gate by electron beam lithography. In the second step,
50 nm deep trenches are etched around this top gate in a SiCl$_4$
plasma. The depletion regions around these trenches define the two
quantum dots. We label them \lq u\rq\ and \lq d\rq, for \lq up\rq\
and \lq down\rq\ in Fig.\ \ref{figure1}. Each quantum dot has its
own source and drain leads, so we can independently measure their
differential conductances ${\rm d}I^{\rm(u)}_{\rm D}/{\rm d}V^{\rm
(u)}_{\rm DS}$ and ${\rm d}I^{\rm(d)}_{\rm D}/{\rm d}V^{\rm
(d)}_{\rm DS}$ \footnote{Here measured by lock-in technique with a
modulation voltage of 1 $\mu$V$_{\rm pp}$.}. Furthermore, each of
the four tunnel barriers is tunable by one of the adjacent gates 1
to 4. It was tested experimentally that the bridge between the two
quantum dots carrying the top gate is entirely depleted, so no
current can flow between the dots and the coupling is purely
capacitive. The top gate is needed to reach a large interdot
capacitance leading to a large ratio between the interdot Coulomb energy $U$
and the intradot charging energies $E_{\rm Cu}$ and $E_{\rm Cd}$
\footnote{Please note, we define $E_{\rm C}=e^2/C_{\Sigma}$, instead
of $E_{\rm C}=e^2/2C_{\Sigma}$.}.

\par
At our base temperature of 25 mK, the conductances at small
source-drain voltages show a honeycomb-like structure as a function
of two gate voltages \cite{Huebel1}, as expected from a simple
electrostatic model \cite{Wiel1}. Fig.~\ref{figure1} gives the basic
definitions to be used for its description. Each possible
combination ($N_{\rm u},N_{\rm d}$) of occupation numbers is stable
inside a hexagonal area in the parameter space spanned by the two
gate voltages. At tunnel couplings $\Gamma_{\rm u}$ and $\Gamma_{\rm
d}$ much smaller than the thermal energy $k_{\rm B}T$, only
single-electron tunneling is possible, and only on those hexagon
edges where exactly one quantum dot changes its occupation number
(edges of type $a$ and $b$). In contrast, no conductance is found on
type $c$ edges where both occupation numbers must change
simultaneously. There, the interdot Coulomb blockade prevents
single-electron tunneling. Near the common pinch-off point of all
four tunnel barriers, we find $U\approx 0.26$ meV, and $U/E_{\rm Cu}\approx U/E_{\rm
Cd}\approx 1/3$. These values decrease as one opens the tunnel
barriers because of increasing dot-lead capacitances \cite{Huebel2}.

\par
As a measure of the tunnel couplings, we take the full widths at
half maximum of the Coulomb peaks on the type $a$ and type $b$
edges. In order to convert these values (measured in units of a gate
voltage) into energies, we determine the capacitive lever arm
between the gate voltage and the quantum dot's addition energy by
finite source-drain voltage measurements as described in
\cite{Huebel2}. The sharp conductance peaks of the weakly tunnel
coupled quantum dots are used in order to precisely evaluate the
energy scale, which is then transferred to the strongly coupled
system. In the experiment, the tunnel couplings are usually large
compared to $k_{\rm B}T$ at the base temperature of 25 mK, which
means that temperature broadening effects are negligible; for type
$a$ or type $b$ edges the conductance peaks show a lineshape which
is approximately Lorentzian in most cases. The interesting regions
we will focus on in the rest of this letter are obviously the edges
of type $c$. Here, we cannot start from the simple single-electron
tunneling picture because of the interdot Coulomb blockade.

\par
\begin{figure}[t]
    \includegraphics[angle=0,width=8.7cm]
      {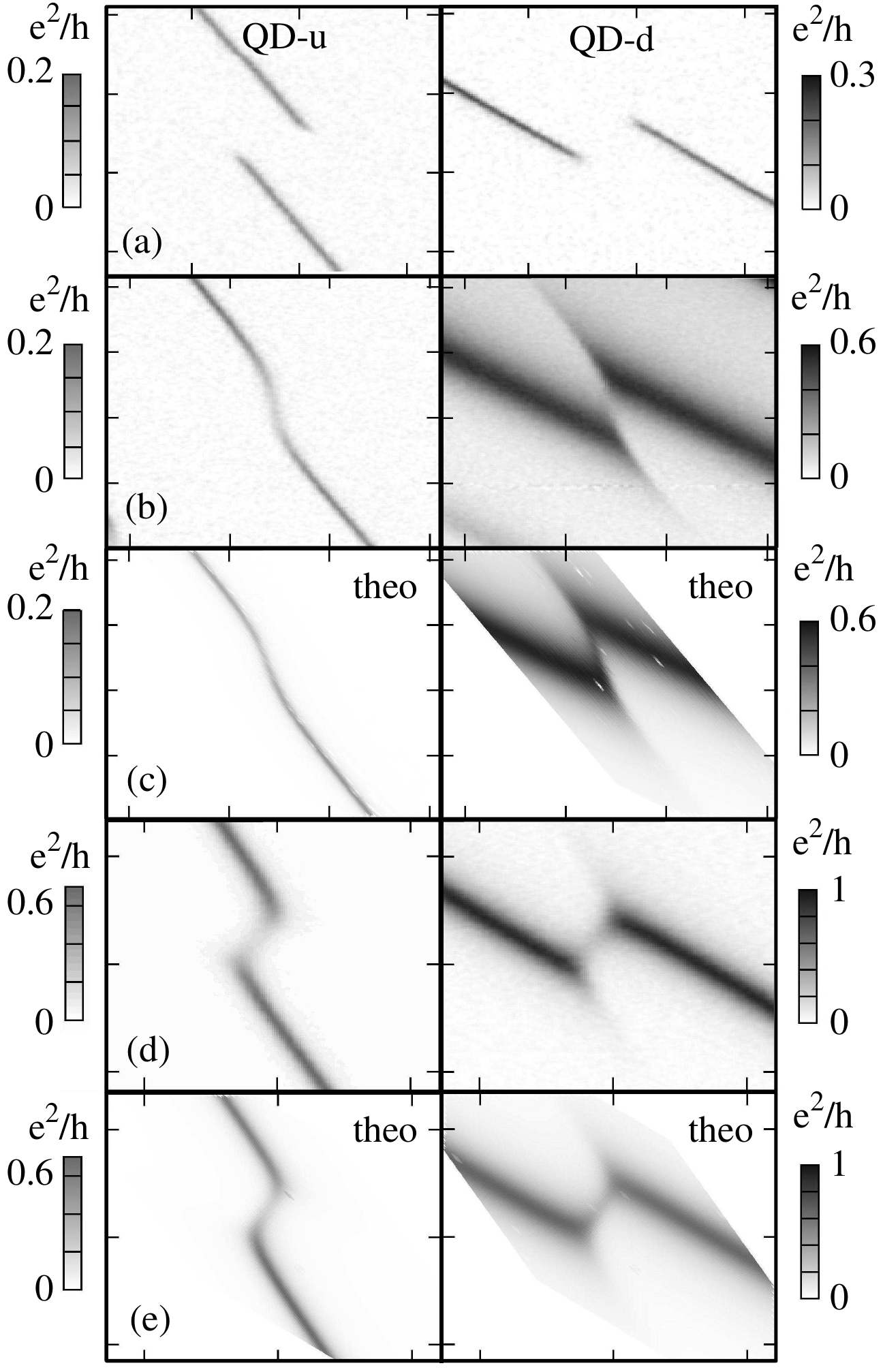}
    \caption{\label{figure2}
    The conductances of quantum dots \lq u\rq\ (left) and  \lq d\rq\
    (right) in greyscale at 25 mK around three type $c$ lines for
    different tunnel couplings. Horizontal axes: $V_{1,2}$; vertical
    axes: $V_{3,4}$ in steps of 2 mV.
    (a),(b),(d) are measured, (c),(e) are calculated by using
    parameters extracted from (b),(d), respectively.
    $\Gamma_{\rm u}=\Gamma_{\rm d}=25$ $\mu$eV
    and $U=260$ $\mu$eV in (a); $\Gamma_{\rm u}=25$ $\mu$eV, $\Gamma_{\rm d}=110$ $\mu$eV,
    and $U=140$ $\mu$eV in (b); $\Gamma_{\rm u}=32$ $\mu$eV, $\Gamma_{\rm d}=59$ $\mu$eV,
    and $U=163$ $\mu$eV in (d).}
\end{figure}

Fig.~\ref{figure2}(a) shows a region with small tunnel couplings in
more detail. Finite conductances can be seen only on the type $a$
and type $b$ edges, so single-electron tunneling provides a
qualitatively sufficient explanation. Fig.~\ref{figure2}(b) shows a
situation in which one of the dots is much more strongly tunnel
coupled to its leads, while the other one remains weakly coupled. A
sharp conductance peak is observed for the weakly coupled dot that
follows a continuous curve, so the interdot Coulomb blockade is
lifted. The peak amplitude is smallest at the turning point of the
position curve, and the lineshape of the conductance peaks was
checked to be Lorentzian for all line cuts along the $V_1$
direction (horizontal). Far from the turning point, the curve becomes straight as
a function of the gate voltages, and we can then take the position
of the conductance peak maximum as a definition of the type $a$
honeycomb edges. A different behavior is observed for the strongly
coupled dot: The conductance plot appears to be divided into two
half planes. On each side, the conductance peaks are simply
described by a Lorentzian whose center defines the type $b$
honeycomb edge. A narrow, step-like transition occurs in between. It
is located at the same position where we observe the conductance
peak in the weakly coupled dot.

\par
For the following discussion, we describe the two dots with an
Anderson impurity model which has a single, spin-degenerate quantum
level in each dot:
\begin{eqnarray}\label{Hamiltonian}
    \hat{H}=&&\!\sum\limits_{i\in\left\{{\rm u,d}\right\}}\!\!\!\big(\varepsilon_i\cdot \hat{n}_i+E_{{\rm C}i}\cdot \hat{n}_{i\uparrow}\hat{n}_{i\downarrow}\big)
    + U\cdot \hat{n}_{\rm u}\hat{n}_{\rm d}\nonumber\\
    &&+\!\!\sum\limits_{Ri,k,\sigma}\!\!\!\varepsilon_k\cdot
    \hat{c}^{\dagger}_{Ri,k,\sigma}\hat{c}^{}_{Ri,k,\sigma}\nonumber\\
    &&+\!\!\sum\limits_{Ri,k,\sigma}\!\!\!\left(t_{Ri}\cdot \hat{a}^{\dagger}_{i,\sigma}\hat{c}^{}_{Ri,k,\sigma}+ {\rm
    h.c.}\right)\;.
\end{eqnarray}
Here, $\hat{a}^{\dagger}_{i,\sigma}$  ($\hat{a}_{i,\sigma}$) create
(annihilate) an electron with spin
$\sigma\in\{\downarrow,\uparrow\}$ in dot $i\in\{{\rm u}, {\rm
d}\}$;
$\hat{n}_{i,\sigma}=\hat{a}^{\dagger}_{i,\sigma}\hat{a}^{}_{i,\sigma}$
and $\hat{n}_i=\hat{n}_{i,\uparrow}+\hat{n}_{i,\downarrow}$, are the
corresponding number operators; $\varepsilon_{\rm u}$ and
$\varepsilon_{\rm d}$ denote the addition energies of the dots
relatively to the source Fermi level, which shift linearly with
applied gate voltages; $\hat{c}^{\dagger}_{Ri,k,\sigma}$ is the
creation operator for an electron in lead $R\in\left\{{\rm
S,D}\right\}$ of system $i\in\left\{{\rm u,d}\right\}$ with
wavenumber $k$, spin $\sigma$ and energy $\varepsilon_k$. Finally,
$t_{Ri}$ denotes the corresponding spin-independent tunnel matrix
element, which translates into a tunnel rate
$\Gamma_i=2\pi\rho\cdot\left({t_{{\rm S}i}}^2+{t_{{\rm
D}i}}^2\right)$ where $\rho$ is  the density of states in the leads.

\par
\begin{figure}[t!]
    \includegraphics[angle=0,width=8.7cm]
      {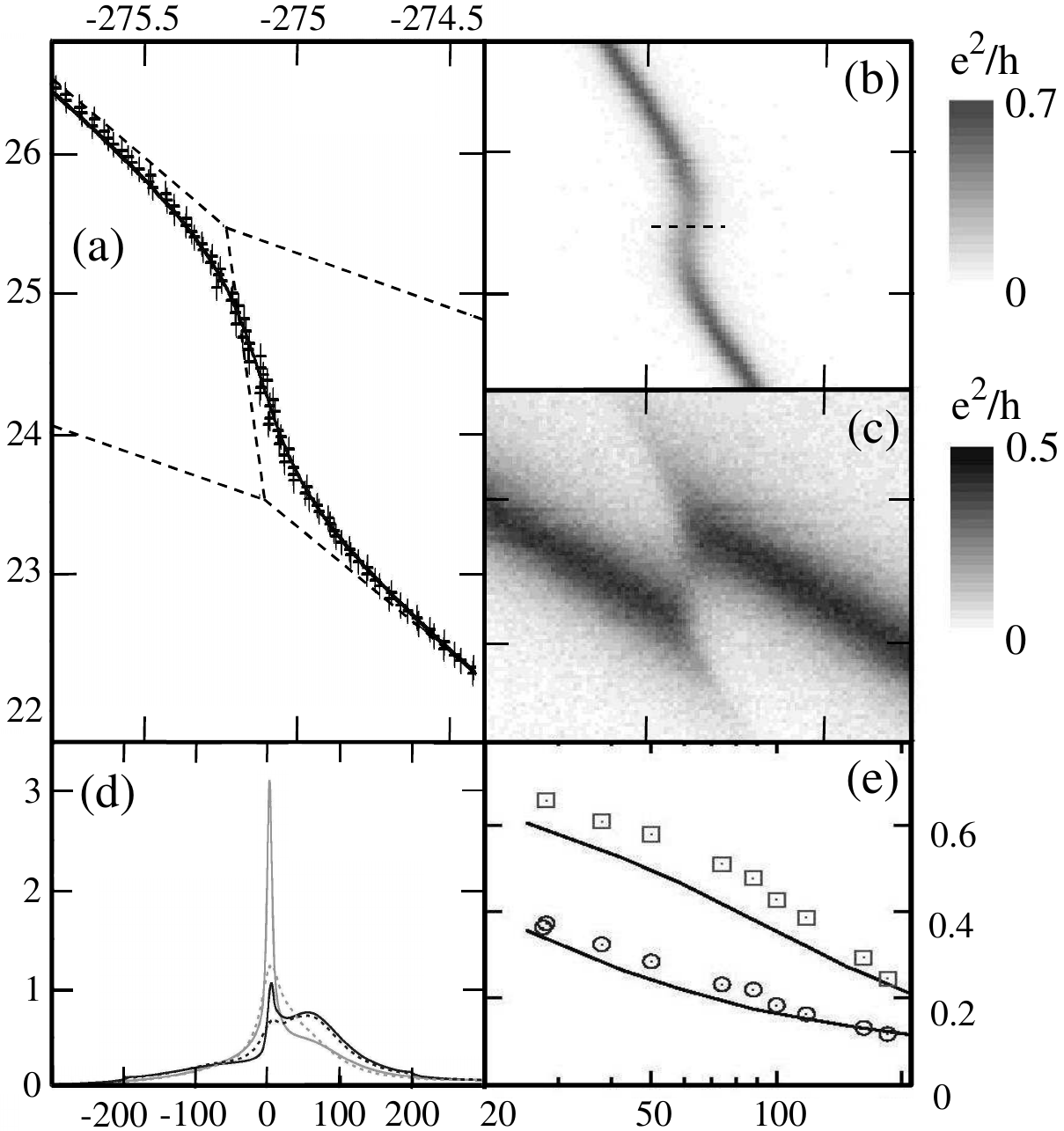}
    \caption{\label{figure3}
    (a) Crosses give the positions of conductance maxima in the upper dot, ex\-trac\-ted from Fig.\ \ref{figure2}(b) (axes in mV).
        Solid line: pre\-dic\-tion of Eq.\ (\ref{curve}), dashed line: ho\-ney\-comb edges.
    (b),(c) Mea\-su\-red con\-duc\-tan\-ces of up\-per and lower dot
    around another type $c$ line at 25 mK (axes like in Fig.\ \ref{figure2}). The measured parameters are $\Gamma_{\rm u}=31$ $\mu$eV,
    $\Gamma_{\rm d}=86$ $\mu$eV, $U=120$ $\mu$eV).
    (d) NRG spec\-tral den\-si\-ties $A(\omega)$ at the middle of the dashed
    line in (b), showing the de\-velop\-ment of a Kondo reso\-nance in both dots
    with de\-crea\-sing tem\-pe\-ra\-ture ($A$ in $\mu$eV$^{-1}$, $\omega$ in $\mu$eV).
    Grey: upper dot, black: lower dot. Solid: $T=25$ mK, dashed: $T=144$ mK.
    (e) Measured peak con\-duc\-tances (in $e^2/h$) for two line cuts of (b) as a func\-tion
    of tem\-pe\-ra\-ture (in mK). Boxes: Cros\-sing the $a$ line at fixed $V_{1,2}$, cir\-cles: cros\-sing the $c$ line at fixed $V_{1,2}$ as shown in (b).
    The solid line give the respective NRG calculation results.}
\end{figure}

To get a simple physical picture, we neglect, in a first
approximation, charge fluctuations in the (much) more weakly coupled dot,
setting $\Gamma_{\rm u}=0$. Since no spin-Kondo effect was observed
in the adjacent Coulomb blockade valleys, and the intradot charging
energies are the largest parameters in the system ($E_{\rm
Cu}\approx E_{\rm Cd}\approx 0.6$ meV in Fig. \ref{figure2}(b,d) and
\ref{figure3}(b,c) obtained from measurements of
the Coulomb blockade diamonds), we further neglect double occupation
of the individual dots and drop the $E_{{\rm C}i}$ terms and the
spin indices in Eq.\ (\ref{Hamiltonian}). The description of the
double-dot ground state then reduces to a resonant tunneling
Hamiltonian for the strongly coupled quantum dot, predicting a
Lorentzian spectral density of width $\Gamma_{\rm d}$ on this dot
\cite{Mahan}. However, the resonant state now has two possible
addition energies, $\varepsilon_{\rm d}$ and $\varepsilon_{\rm
d}+U$, depending on the (fixed) occupation number $N_{\rm
u}\in\left\{0,1\right\}$ of the weakly tunnel coupled dot. Like in
the electrostatic model \cite{Wiel1}, we assume that the system
takes the state of minimal electrostatic energy $E$, which is given
by the expectation value of $\varepsilon_{\rm u}\cdot \hat{n}_{\rm
u}+\varepsilon_{\rm d}\cdot \hat{n}_{\rm d}+U\cdot \hat{n}_{\rm u}
\hat{n}_{\rm d}$:
\begin{eqnarray}
    \nonumber
    E(N_{\rm u})=&&\varepsilon_{\rm u}\cdot N_{\rm u}+\Big(\varepsilon_{\rm d}+
    U\cdot N_{\rm u}\Big)\\
    &&\times\left[\frac{1}{2}-\displaystyle\frac{1}{\pi}\cdot\arctan
    \displaystyle\frac{\varepsilon_{\rm d}+U\cdot N_{\rm u}}{\Gamma_{\rm d}/2}\right]\;.
\end{eqnarray}
Conductance through the weakly coupled dot is now possible for
$E(N_{\rm u}=1)=E(0)$, even if charge fluctuations ($\Gamma_{\rm
u}$) are very small. This yields a relation between
$\varepsilon_{\rm d}$ and $\varepsilon_{\rm u}$, which we express
after substituting $\varepsilon_i$ by $\varepsilon_i'-U/2$ as
\begin{eqnarray}\label{curve}
    \nonumber
    \varepsilon_{\rm u}'=&&\Big(\varepsilon_{\rm d}'+U/2\Big)\cdot\displaystyle\frac{1}{\pi}
    \arctan{\displaystyle\frac{\varepsilon'_{\rm d}+U/2}{\Gamma_{\rm d}/2}}\\
    &&-\Big(\varepsilon_{\rm d}'-U/2\Big)\cdot\displaystyle\frac{1}{\pi}
    \arctan\frac{\varepsilon_{\rm d}'-U/2}{\Gamma_d/2}\;.
\end{eqnarray}
Note the point symmetry around $\varepsilon_{\rm
d}'=\varepsilon_{\rm u}'=0$. Only the ratios
$\varepsilon_i'/U$ and $\Gamma_{\rm d}/U$ enter, which can be measured to
within a few per cent. The two quantities $\varepsilon_i'/U$ are
linear functions of the gate vol\-ta\-ges entirely determined by
capacitance ratios, i.e.\ the slopes of the honeycomb edges.
By evaluating Fig.~\ref{figure2}(b), we get the transformation
relation
\begin{eqnarray}\label{transformation}
\left(\!\!\begin{array}{c}{V_1-V_1^{(0)}}\\
{V_3-V_3^{(0)}}\end{array}\!\!\right)\!/{\rm mV} =
\left(\!\!\begin{array}{rr} -1.62(0)& 1.74(3) \\
1.26(0)&-3.19(8) \end{array}\!\!\right)\left(\!\!\begin{array}{c} \varepsilon_{\rm u}'/U\\
\varepsilon_{\rm d}'/U\end{array}\!\!\right).
\end{eqnarray}
The prediction of Eqs.~(\ref{curve}) and (\ref{transformation}) is
plotted in Fig.~\ref{figure3}(a). $(V_1^{(0)},V_3^{(0)})$
describes the turning point of the position curve for the
conductance peak in the gate voltage plane. It should be at the
center of the type $c$ line, which is constructed by
extrapolating the positions of the (single-electron like)
$a$ and $b$ lines into the regions of interaction. In
Fig.~\ref{figure3}(a) we see that the turning point
$(V_1^{(0)},V_3^{(0)})$ is slightly offset by $0.13$ mV in $V_1$ (horizontally) and
$-0.05$ mV in $V_3$ (vertically). This deviation indicates that the assumption of
a Lorentzian spectral density for the strongly tunnel coupled dot
simply being shifted by recharging the weakly coupled dot is not
strictly fulfilled. Such an offset could be caused, for example, by
an asymmetric spectral density and/or energy dependent tunnel
barriers. However, apart from this small offset, the calculated
curve reproduces the position shift of the conductance peak with
good accuracy.

\par

Predicting the value of the peak conductance is a much more
difficult task. It varies along the position curve, reaching a
minimum at the turning point. With increasing temperature,
conductance decreases on all parts of the curve, notably at
the turning point (Fig. \ref{figure3}(e)), so it is essential to calculate at
finite temperature. To this end, we performed NRG
calculations \cite{NRG} for Hamiltonian (\ref{Hamiltonian}) with
experimentally predetermined parameters and taking into account a
reduction of the tunnel couplings $\Gamma_{\rm u}$, $\Gamma_{\rm d}$ by a factor of 2 because
many-body effects lead to an additional broadening of the side bands
of the Anderson impurity model \cite{holedecay}. The NRG results for the
region around the $(N_{\rm u},N_{\rm d})=(0,1)/(1,0)$ boundary in
Fig. \ref{figure2} (c) and (e) agree well with experiment,
including a reduced conductance around the turning points. Near the
turning points of Fig. \ref{figure2}(e) and Fig. \ref{figure3}(b), an
additional Kondo resonance develops at the Fermi level in
the NRG spectra, see Fig. \ref{figure3}(d). Hence, for tunnel couplings
$\Gamma_{\rm u}$, $\Gamma_{\rm d}$ both tuned to intermediate values, we can trace back the
conductance along the type $c$ line to (orbital) Kondoesque tunnel
processes still above the Kondo temperature.  We also note two
deviations if we restrain ourselves from adjusting the parameters:
For the upper dot, the theoretical curve is somewhat smoother in
the region of the type $c$ line; for the lower dot the conductance
peak height is lower since many-body effects lead to a transfer of
spectral weight so that the behavior is more complicated than the
simple shift of the Lorentzian-broadened peak assumed when
determining the tunnel rate. Both deviations could be reduced if the
tunnel rate to the lower dot was reduced. Also effects beyond our
calculation, in particular decoherence and the importance of more
than one level in the lower dot, would lead to an adjustment in the
same direction.

In conclusion, having full control over tunnel couplings and gate
voltages of two laterally arranged quantum dots and measuring the
conductance through the two dots separately, we were able to
unambiguously identify regions of single-electron tunneling and
correlated Kondoesque tunneling. Our experimental and theoretical
analysis shows that the interdot Kondo effect leads to
conductance through both dots in the region of the type $c$ line if
tunnel couplings are roughly symmetrical albeit the temperature
is still sightly above the Kondo temperature.

We acknowledge helpful discussions with J.\ Bauer and Th.\ Pruschke,
and thank Th.\ Pruschke for making available his NRG
program. The project has been supported by the BMBF under grant
01BM455 and the DFG within the SFB/TRR21.

\end{document}